\RequirePackage{fix-cm}
\documentclass[twocolumn]{svjour3}
\pdfoutput=1
\usepackage{graphicx}
\usepackage{amsmath}
\usepackage{amsfonts}
\usepackage{amssymb}
\usepackage{bm}
\usepackage{color}
\usepackage[]{algorithm2e}

\newcommand*\justify{%
	\hyphenchar\font=`\-
}

\newcommand{\unitVec}[1]{\hat{\bm{#1}}}

\begin{document}

\title{Stable algorithm for event detection in event-driven particle
  dynamics: logical states}

\author{Severin Strobl \and Marcus N. Bannerman \and Thorsten P\"oschel}

\institute{
  Severin Strobl \and Thorsten P\"oschel \at
  Institute for Multiscale Simulation,
  Friedrich-Alexander-Universit\"at Erlangen-N\"urnberg,
  N\"agelsbachstr. 49b, 91052 Erlangen,
  Germany\\
  \email{severin.strobl@fau.de}
  \and
  Marcus N. Bannerman \at
  School of Engineering,
  University of Aberdeen,
  Fraser Noble Building,
  Aberdeen,
  AB24 3UE, UK
}

\date{Received: date / Accepted: date}

\maketitle

\begin{abstract}
	Following the recent development of a stable event-detection
	algorithm for hard-sphere systems, the implications of more complex
	interaction models are examined. The relative location of particles
	leads to ambiguity when it is used to determine the interaction
	state of a particle in stepped potentials, such as the square-well
	model. To correctly predict the next event in these systems, the
	concept of an additional state that is tracked separately from the
	particle position is introduced and integrated into the stable
	algorithm for event detection.
\keywords{DEM; Event-driven; Molecular dynamics; Square well; Stepped
    potential; Collision detection;}
\end{abstract}


\section{Introduction\label{sec:introduction}}
\begin{figure}
  \begin{center}
    \includegraphics[width=\columnwidth,clip]{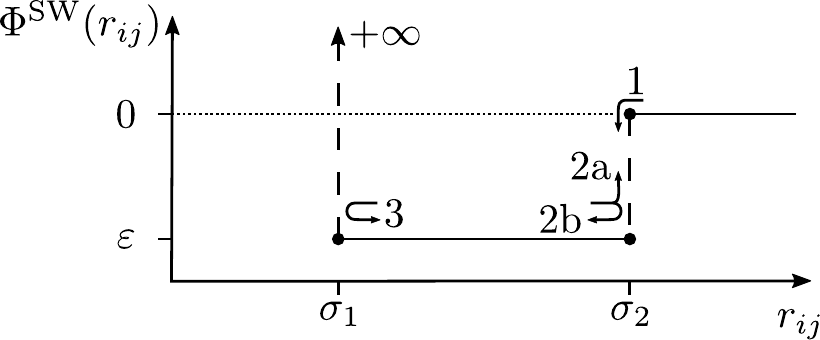}
  \end{center}
  \caption{\label{fig:square-well}%
	The potential energy $\phi^\text{SW}\left(r_{ij}\right)$ of the square-well
	model as a function of the separation distance, $r_{ij} = |\vec{r}_{ij}|
	\equiv |\vec{r}_i - \vec{r}_j|$, between two particles $i$ and $j$. The
	inter-particle energy only changes in discrete steps at distances of
	$r_{ij}=\sigma_1$ and $r_{ij}=\sigma_2$. The numbered arrows indicate the
	four event types of the model generated by these steps: (1) Capture (enter
	the well), (2a) Release (escape the well), (2b) Bounce (remain in the
	well), and (3) Core (hard-sphere collision).}
\end{figure}

Particle dynamics is the numerical solution for the motion of a
collection of discrete bodies, each of which may represent objects
ranging in size from atoms/molecules (molecular dynamics) to grains of
sand or the ice in an avalanche (granular dynamics). There is a variety
of particle dynamics approaches but common to all is the integration
of Newton's equation of motion to determine the trajectory of the
particles. Event-driven particle dynamics (EDPD) is one approach which
integrates Newton's equation of motion exactly through piece-wise
analytic solutions of the particle trajectories. This avoids
truncation error which arises if a numerical integration technique is
used, but restricts the simulation to interactions where piece-wise
analytic solutions can be found. One such class of compatible
interactions is discrete potentials, such as the square-well model
shown in Fig.~\ref{fig:square-well}. As there are no forces acting
between discontinuities in the potential, the analytical solution to
the dynamics is a simple ballistic motion of the particles. When
particles cross a discontinuity, the instantaneous energy change
results in an impulse and this {\em event} must be detected and
processed at the exact time it occurs. The time of the next event is
calculated {\em a priori} and the system is advanced forward in time
directly to the instant of the event: i.e., the progression of time is
event-driven. This allows EDPD implementations to be computationally
efficient, particularly for dilute or stiff systems, as the
uninteresting time between interactions and hence events is skipped.

In a recent paper~\cite{Bannerman_etal_2014} we
demonstrated that the EDPD algorithm must be carefully constructed to
ensure its numerical stability. Although the EDPD algorithm is exact,
its implementation is sensitive to round-off error. In hard-sphere
systems, round-off error manifests as overlapping/interpenetrating
particles which is difficult to resolve as the dynamics is undefined
in this state. If particles begin to interpenetrate, the stable
algorithm executes additional events to ensure the interpenetrating
particles do not continue to approach and, thus, increase their
overlap~\cite{Bannerman_etal_2014}. Key to the stable algorithm
is the definition of valid and invalid states but this distinction is only
straightforward for hard interactions.
For example, particles interacting via a square-well potential (see
Fig.~\ref{fig:square-well}) may overlap ($r_{ij}<\sigma_1$) which is
always an invalid state, but they can also be either in a {\em captured}
state within ($\sigma_1\le r_{ij}\le\sigma_2$) or an {\em uncaptured}
state outside of the well ($r_{ij}\ge\sigma_2$).
As a particle pair cannot energetically occupy both states, only one
region is valid at a particular point in their trajectory. Thus, for
stepped or multi-state interactions such as the square-well or stepped
Lennard-Jones potentials~\cite{Thomson_etal_2014}, the valid state
dynamically changes with time.

In this paper, the stable algorithm of Ref.~\cite{Bannerman_etal_2014} is
extended to interactions with dynamically changing valid states. In
Sec.~\ref{sec:ambiguous_states}, the numerically calculated position is
demonstrated to be an unreliable indicator for the current state of the
particle and additional logical state tracking is recommended. The
extended stable algorithm for square wells is then outlined in
Sec.~\ref{sec:square-well}. In Sec.~\ref{sec:virtual_states}, it is
demonstrated that states must be tracked even for virtual/zero-impulse events.
Finally, the tracking of states is validated for two example
configurations in Sec.~\ref{sec:validation}, illustrating the challenges with
conventional approaches, before the conclusions are drawn in
Sec.~\ref{sec:conclusions}.

\section{Ambiguity of particle state\label{sec:ambiguous_states}}

When simulating systems with multiple valid states, it is crucial that
the current valid state can be determined during the simulation in a
reliable manner. The simplest approach is to try to use
the position of the particles to compute the current state.
Unfortunately, calculating the valid state this way is not always
unambiguous.
The difficulty stems from the fact that the captured state, $r_{ij} \in
\left[\sigma_1, \sigma_2\right]$, and uncaptured state, $r_{ij} \in
\left[\sigma_2, \infty\right)$, are both closed sets which include the
point $r_{ij}=\sigma_2$. This is required as $r_{ij}=\sigma_2$ is the
point of transition between the two states. As events are instantaneous,
there can be no change in position during the transition.
The ambiguity of particle state is thus not a numerical artifact, but is
always encountered even in a precise simulation devoid of the
peculiarities of floating-point calculations.
To illustrate this further, consider the basic algorithm for the
simulation of particles interacting via a square-well
potential~\cite{Alder_Wainwright_1959}:

\begin{enumerate}
  \item For each particle pair:\label{step:detect}
    \begin{enumerate}
    \item If they are uncaptured:\\
      Test for capture events (type 1, compare
      Fig.~\ref{fig:square-well}).
    \item If they are captured:\\
      Test for bounce/release (type 2) and core (type 3) events.
    \end{enumerate}
  \item Sort the events to determine which one occurs first.
  \item \label{step:stream} Move the system forward to the time of the
    first event, $t+\Delta t$.
  \item For the interacting particle pair:\label{step:interact}
    \begin{enumerate}
    \item If they are uncaptured:\\
      Execute a capture event (type 1).
    \item If they are captured:\\
      Execute a bounce/release (type 2) or core (type 3) event.
    \end{enumerate}
  \item Return to step~\ref{step:detect}.
\end{enumerate}

\begin{figure}[h]
  \begin{center}
    \includegraphics[width=\columnwidth,clip]{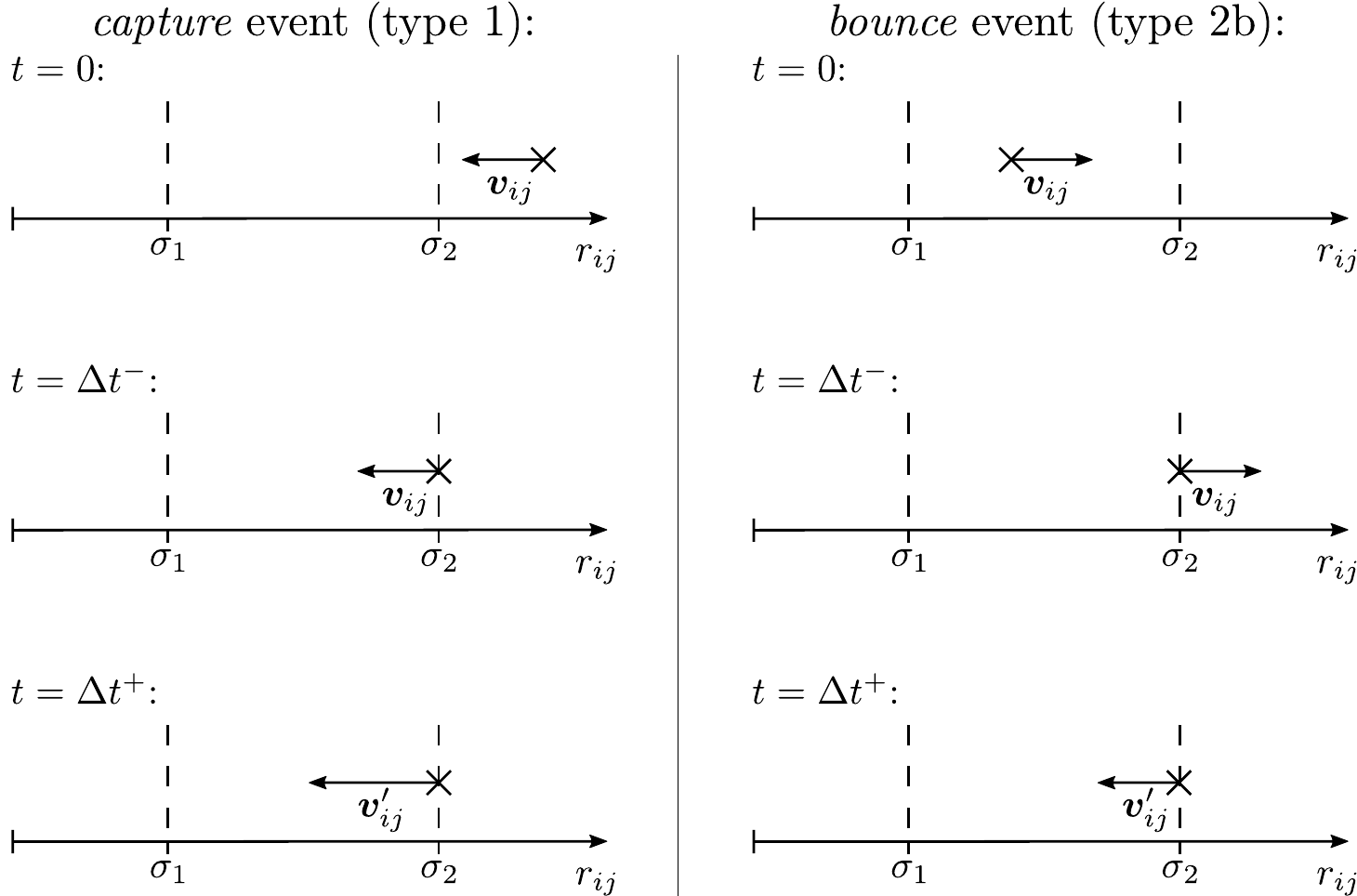}
  \end{center}
  \caption{\label{fig:ambiguity}%
	Two configurations of a pair of particles $i$ and $j$ interacting
	via a square-well potential which lead to an ambiguity in the
	interaction state. The relative velocity $\vec{v}_{ij} \equiv
	\vec{v}_i - \vec{v}_j$ and its post-event value $\vec{v}_{ij}'$ are
	indicated by the arrows. In the left case, uncaptured particles
	enter the well of the potential (event type 1 in
	Fig.~\ref{fig:square-well}) and their capture state changes. In the
	right case, two captured particles bounce on the well discontinuity
	(event type 2b in Fig.~\ref{fig:square-well}) and do not change
	their capture state. In both cases, the position before ($t=\Delta
	t^-$) and after ($t=\Delta t^+$) the event is identical, thus it
	cannot be used to determine the capture state immediately before or
	after an event.}
\end{figure}

In steps 1 and 4, the capture state of the particles must be
determined. Consider the case where the next event to occur arises
from the $r_{ij}=\sigma_2$ discontinuity and is either a capture or a
release/bounce event. The particles will be moved in
step~\ref{step:stream} to the moment of interaction and will lie
exactly on the well edge at $r_{ij}=\sigma_2$. At this
point, the position has become ambiguous for testing the particle state
and this failure occurs for all such events.
In Fig.~\ref{fig:ambiguity} two examples of square-well particles
interacting via different event types are sketched. Independent of the
original state (uncaptured or captured), during the event both
particles have a separation distance $r_{ij}=\sigma_2$ leading to the
ambiguity in the type of the event to execute.

In fact, in real implementations the ambiguity is not limited to a
single point in the separation distance, but expands to an interval
around the discontinuity due to the limited precision of floating-point
calculations. This is illustrated for hard spheres in Fig.~3 of
Ref.~\cite{Bannerman_etal_2014} where particles moved during a core
event have a distribution of separation distances, either slightly
smaller or larger than the expected value. In square-well systems, this
implies that particles moved into position for a well event will end up
distributed both inside and outside of the well, regardless of which
side they are on originally. While it might seem tempting to resolve
the ambiguity at the discontinuity by taking into account the sign of
the relative velocity in addition to the position of the particles, the
distribution of the separation distance on both sides of the
discontinuity impedes this approach.

The first published EDPD implementation by Alder and
Wainwright~\cite{Alder_Wainwright_1959} reduces the effects of this
inaccuracy by storing the separation distance, $r_{ij}$, calculated in
step 1 and uses it in step 4 to determine the capture state. As this
separation is calculated before the particles are moved into place for
the event, there is a lower probability of failure; however, there is
no inherent guarantee in this approach that the stored position
calculated in step 1 is always unambiguous and free from error. The
dynamics do not exclude particles from being located at or near a
discontinuity during event detection. In fact, a particle which has
just executed an event {\em must} be located near a discontinuity and
ambiguity must arise immediately in the next iteration of step
1. Alder and Wainwright do not specify how event detection is carried
out in this case~\cite{Alder_Wainwright_1959}.

There are a number of approaches in the literature which try to
resolve the ambiguity arising from particles near to a discontinuity.
One approach is to perform ``corrections'' to ensure that particles
numerically transfer to the correct side of the discontinuity. This
may be achieved through the addition or subtraction of a small
quantity to the time of the next
event~\cite{Poschel_Schwager_2005}. Alternatively, the position may be
directly changed to ``nudge'' the particles over the corresponding
discontinuity after processing the current event (e.g., see the source
code of Ref.~\cite{Schultz_Kofke_2015}). Unfortunately, both methods
rely on empirically determined correction factors, which depend on
the system studied and may introduce other errors if another event
occurs at around the same time. For example, if a third particle is in
close proximity then stretching the event time may cause an
interaction to be missed.

The infinitely-thin hard rods system~\cite{Frenkel_Maguire_1983} is
particularly interesting as it naturally exhibits a large number of
repeated re-collisions between pairs of particles which have just
collided. To prevent spurious re-detection of collisions which have
just been executed, a minimum re-collision time computed from the
underlying dynamics is enforced. This requires the storage of the last
event of each particle to correctly apply the minimum re-collision
time restriction. This approach is highly appealing as it uses some
additional {\em logical} (not floating point) state to enforce that the
correct system dynamics is generated.

It appears that, for systems where the valid states change with time,
the simplest robust approach is to explicitly track the current state
of particle pairs as additional logical information within the
system. In square-well systems, the required logical state is boolean
(``captured'' or ``uncaptured'') but in multi-step potentials there
may be many logical states.

Initialization of the logical state may take place from the positional
information only when the configuration is first generated (using an
arbitrary choice to resolve the $r_{ij}=\sigma_2$ case). For all later
times, the logical state must be stored and loaded along with the rest
of the configurational information whenever the simulation is suspended
or resumed. As the logical state only changes when the correct
``transfer'' event is executed, and only logically consistent events are
tested, all ambiguity is eliminated. The stable implementation for
square-well molecules is now outlined.

\section{Stable event detection in square wells\label{sec:square-well}}

The introduction of logical states requires some modifications to the
stable algorithm for event detection as outlined in
Ref.~\cite{Bannerman_etal_2014}. At the core of the stable algorithm
lies the definition of an overlap function, $f(t)$, which indicates
whether a pair of particles is in a valid state or not at the time $t$.
The square-well system is composed of multiple applications of the
overlap function $f_\text{BB}$ for two closed balls, which has the
following form:
\begin{align}
  \label{eq:bb-overlapfunc}
  f_\text{BB}(t + \Delta t, \sigma) = \left[\vec{r}_{ij}(t) + \Delta t\,\vec{v}_{ij}(t)\right]^{2} -\sigma^2\,,
\end{align}
where $\vec{v}_{ij}(t) \equiv \vec{v}_i(t) - \vec{v}_j(t)$ is the
relative velocity, and $\sigma$ is the average diameter of the two
balls. This function is negative if the balls are overlapping, positive
if they are apart, and zero if they are in contact. Thus event detection
for the intersection of two closed balls is transformed into the
solution for the roots of $f_\text{BB}$.

The time derivative of an overlap function $\dot{f}(t)$ can be used to
distinguish whether a currently invalid state ($f(t) < 0$) is either
improving or stable in time ($\dot{f}(t)\ge0$) or deteriorating
($\dot{f}(t)<0$). For $f_\text{BB}$ this corresponds to
\begin{align}
  \label{eq:bb-overlapfunc-deriv}
  \dot{f}_\text{BB}(t + \Delta t) = \vec{v}_{ij}(t)\cdot\left[\vec{r}_{ij}(t) + \Delta t\,\vec{v}_{ij}(t)\right]\,.
\end{align}
Following the algorithm in Ref.~\cite{Bannerman_etal_2014}, a
stable EDPD algorithm is defined as:

{\em When testing for interactions between a pair of particles at a
  time, $t$, consider all logically valid overlap functions. For each
  overlap function, $f$, an event occurs after the smallest
  non-negative time interval, $\Delta t$, that satisfies the following
  condition:}
\begin{align}
  \left(\vphantom{\dot{f}} f\left(t+\Delta t\right)\le 0\right) \text{{ }and }
    \left(\dot{f}\left(t+\Delta t\right)<0\right)\,.
\end{align}
This algorithm prevents errors in positional state from deteriorating
once they are detected. If the two particles are uncaptured then only
capture events are logically valid and are tested for using
$f_\text{BB}(t,\,\sigma_2)$. If the two particles are already
captured, then core events are tested for using
$f_\text{BB}(t,\,\sigma_1)$ and release/bounce events must also be
tested for using $-f_\text{BB}(t,\,\sigma_2)$. The negative of an
overlap function is its inverse which results in $-f_\text{BB}$
becoming a test for when the particles will exceed the well distance
$\sigma_2$ and become disjoint. Both core and bounce/release events are
tested for at the same time and the earliest event is taken as the next
event. As these functions represent distinct invalid state volumes in
$\bm{r}_{ij}$, there is no chance of ambiguity through simultaneous
events from both overlap functions. A reference implementation of the
event-detection algorithm using the logical state and the corresponding
overlap functions is given in Appendix~\ref{app:ref-impl}.

\section{\label{sec:virtual_states}Virtual states}

In Sec.~\ref{sec:ambiguous_states}, logical states are introduced for
systems with dynamic valid states and applied in the context of
discrete potentials. In these systems, the different logical states
have a direct relationship to potential energy changes in the
interaction potential; however, there are many cases where additional
state tracking is desirable even though it is not associated with any
physical action or impulse.
These states may hence be considered as {\em virtual} states
as they have no physical meaning.
One example is neighbor-lists where the simulation domain is divided
into small subvolumes or neighborhoods to optimise the search for the
next event~\cite{Rapaport_1980,Marin_etal_1993}. Each neighborhood
corresponds to a different virtual state for each particle, and only
particle pairs in neighboring states are actually tested for events. It
is vital that, even in these virtual state systems, the logical state is
tracked and used to guarantee the correct dynamics.

The simplest example of this is a particle simulation where a virtual wall
divides the simulation domain into two half-spaces. Virtual walls may be used
to track transport properties such as the mass flux of particles across the
wall, as well as form part of a neighbor-list implementation. The wall is
virtual as it is merely a bookkeeping device and is penetrable
by all particles.
The wall divides the simulation space into two closed half-spaces. A
suitable overlap function for one of these half-spaces is
\begin{align}
  f_\text{HS}(t+\Delta t) = \unitVec{n}\cdot\left[\vec{r}_{i}(t+\Delta t) - \vec{c}\right]\,,
\end{align}
where the half-space is defined by a normal $\unitVec{n}$
perpendicular to the virtual wall and a point $\vec{c}$ in the plane.
The direction of the normal depends on which side of the wall the
particle is currently on. As the overlap function is required to be
positive for valid states, the normal must point into the half-space
the particle is currently located in. If a particle is located either exactly or
numerically close to the virtual wall it is unreliable to determine the
current half-space (and hence the direction of the normal) from the
position. Therefore the side of the wall the particle is
located on must be tracked as additional logical state.


\section{Validation\label{sec:validation}}

To illustrate the significance of state tracking, simulations of two
model systems are performed.  The first model is a square-well
potential (Fig.~\ref{fig:square-well}) with parameters
$\sigma_2 = 3\,\sigma_1$, and $\varepsilon = -k_B\,T$. The second
model is an equivalent square-shoulder potential where
$\varepsilon = +k_B\,T$. These potentials are the prototypes of more
complex interaction models such as general stepped
potentials~\cite{Thomson_etal_2014}. Both simulations use
$N = 32\,000$ particles and a reduced number density of
$N\,\sigma_1^3/V=10^{-3}$. Simulations are run for 10\,000 events per particle
and an instrumented version of the DynamO \cite{Bannerman_etal_2011} simulation
package is used to collect statistics on inconsistencies which arise in the
logical and physical state.

For the square-well system, the probability of inconsistencies
arising during event processing is approximately $1.4\times 10^{-7}$
for each event, while for the square-shoulder system it is
approximately $2.6 \times 10^{-7}$. Although these events occur
relatively infrequently the results can be catastrophic. The
implementation of Alder and Wainwright~\cite{Alder_Wainwright_1959}
becomes stuck and repeatedly executes identical release events (type 2a in
Fig.~\ref{fig:square-well}). The simulation cannot proceed forward in time and
the kinetic energy increases with each event. The logical states algorithm
proposed here is able to continue the simulation and maintains the total energy
to machine precision.

The rare nature of the inconsistency explains why it was not reported
earlier in the literature; however, simulations of $10^{10}$ events or
more (e.g., see Ref.~\cite{Bannerman_Lue_2010a}) are now commonplace,
thus care must be taken to ensure the simulation remains
unconditionally stable.

\section{Conclusions\label{sec:conclusions}}

In conclusion, the physical and virtual state of an EDPD simulation
must not be derived from the position of the particles except
during initialisation. This information must instead be tracked as
additional logical state and used to enforce that the correct event
sequences are both detected and executed. The logical states represent
a crucial part of the state of the system as a whole so, if the
simulation is suspended, all logical states must be stored and
restored accordingly once the simulation is restarted. During the
runtime of the EDPD simulation the logical states must only evolve via
the execution of the correct transfer events. For example, for the
previously examined square-well model, only the event types 1 and 2a
(Fig.~\ref{fig:square-well}) for particles entering or leaving the
square well lead to a change in the logical state for this pairing of
particles.

Unlike the configurational state (which is inherent to each particle in
the system), logical state may be required for each possible pairing of
interacting particles. Thus the number of stored states may scale as
$\mathcal{O}(N^2)$ in the number of particles, $N$. To avoid
difficulties in scaling to large systems, efficient hashed
implementations are recommended where only variations from the most
common state are stored. Typically, this reduces the required storage
to $\mathcal{O}(N)$.

If the idea of logical states is combined with the stable algorithm
for event detection, inherently stable EDPD simulations are
possible. This is shown in Sec.~\ref{sec:validation} for simulations
of systems using square-well or square-shoulder
potentials. Furthermore simulations of more complex interaction models
without any artificial modification of the list of predicted events or
interference with the dynamics of the system are enabled.

\begin{acknowledgements}
The authors gratefully acknowledge the support of the German Research
Foundation (DFG) through the Cluster of Excellence 'Engineering of Advanced
Materials' at the University of Erlangen-Nuremberg and through Grant
Po~472/25.
\end{acknowledgements}

\appendix

\section{\label{app:ref-impl}Stable algorithm for square-well molecules}

The calculation of the event times in Sec.~\ref{sec:square-well} is
expressed in terms of the ball--ball overlap function, $f_\text{BB}$, as
given in Eq.~\ref{eq:bb-overlapfunc}. This operation is especially
sensitive to round-off error in the floating-point representation,
therefore two numerically robust subroutines which analyse $f_\text{BB}$
are presented here. Both use the quadratic equation to solve for the
roots of $f_\text{BB}$; however, each has different safeguards against
numerical errors. The first algorithm, \texttt{\justify
BallBallIntersectionTime} (Algorithm~\ref{alg:BallBallIntersection}),
calculates the time until two balls begin to intersect. The second,
\texttt{\justify BallBallDisjointTime}
(Algorithm~\ref{alg:BallBallDisjoint}), calculates the time until two
balls become disjoint. Both subroutines return $+\infty$ if no event is
detected and use the appropriate numerically stable form of the
quadratic equation.
\begin{algorithm}[ht]
  \DontPrintSemicolon
  \SetKw{Fn}{Function}{}
  \Fn{{\rm BallBallIntersectionTime}($\vec{r}_{ij}$, $\vec{v}_{ij}$, $\sigma$)}\;
    $a\longleftarrow \vec{v}_{ij}\cdot\vec{v}_{ij}$\;
    $b\longleftarrow \vec{r}_{ij}\cdot\vec{v}_{ij}$\;
    $c\longleftarrow \vec{r}_{ij}\cdot\vec{r}_{ij} - \sigma^2$\;
    $arg\longleftarrow b^2 - a\,c$\;
    \lIf{$b\ge0$ {\rm\bf or} $arg\le0$}{\Return $+\infty$}
    \Return max($0$, $c\,/\,(\sqrt{arg} - b)$)\;
    \vspace{.1cm}
    \caption{\label{alg:BallBallIntersection} A stable algorithm for
      detection when two initially disjoint balls begin to
      intersect. In terms of the overlap function, this determines the
      time until $f_{BB}\le0$ and $\dot{f}_{BB}<0$ or returns
      $+\infty$ if this does not occur in the future.}
\end{algorithm}
\begin{algorithm}[ht]
  \DontPrintSemicolon
  \SetKw{Fn}{Function}{}
  \Fn{{\rm BallBallDisjointTime}($\vec{r}_{ij}$, $\vec{v}_{ij}$, $\sigma$)}\;
    $a\longleftarrow \vec{v}_{ij}\cdot\vec{v}_{ij}$\;
    $b\longleftarrow \vec{r}_{ij}\cdot\vec{v}_{ij}$\;
    $c\longleftarrow \vec{r}_{ij}\cdot\vec{r}_{ij}-\sigma^2$\;
    $arg\longleftarrow b^2 - a\,c$\;
    \lIf{$a=0$}{ \Return $+\infty$}
    \lIf{$arg\le0$}{ \Return max($0$,\,$-b\,/\,a$)}
    \eIf{$b > 0$}{
      \Return max($0$, $(\sqrt{arg}-b)\,/\,a$)\;
    }{
      \Return max($0$, $-c\,/\,(\sqrt{arg} + b)$)\;
    }
    \vspace{.1cm}
    \caption{\label{alg:BallBallDisjoint} A stable algorithm for
      detecting when two initially intersecting balls become
      disjoint. This determines the time until $f_{BB}\ge0$ and
      $\dot{f}_{BB}>0$ or returns $+\infty$ if this does not occur in
      the future.}
\end{algorithm}

The introduction of the logical state and the overlap function requires
some changes to the detection of events as outlined in
step~\ref{step:detect} of the basic algorithm given in
Sec.~\ref{sec:ambiguous_states}. The modified algorithm is presented in
\texttt{\justify SWEventTime} (Algorithm~\ref{alg:square-well}). The
logical state is required as input to this function and must be tracked
seperately. For the square-well model, this is a single Boolean value
per particle pair indicating whether the particles are captured or not.
The specialized routines of Algorithms \ref{alg:BallBallIntersection}
and \ref{alg:BallBallDisjoint} are then used to determine the roots of
the overlap function. In the case of a captured particle pair, both
discontinuities at $\sigma_1$ and $\sigma_2$ are accessible and the
minimum of the respective event times has to be selected.
\begin{algorithm}[ht]
  \DontPrintSemicolon
  \SetKw{Fn}{Function}{}
  \Fn{{\rm SWEventTime}($\vec{r}_{ij}$, $\vec{v}_{ij}$, $\sigma_1$, $\sigma_2$, captured)}\;%
    \eIf{\rm captured}{
      $\Delta t_1\longleftarrow$BallBallIntersectionTime($\vec{r}_{ij}$, $\vec{v}_{ij}$, $\sigma_1$)\;
      $\Delta t_2\longleftarrow$BallBallDisjointTime($\vec{r}_{ij}$, $\vec{v}_{ij}$, $\sigma_2$)\;
      \Return min($\Delta t_1$, $\Delta t_2$)
    }{
      \Return BallBallIntersectionTime($\vec{r}_{ij}$, $\vec{v}_{ij}$, $\sigma_2$)\;
    }
  \vspace{.1cm}
  \caption{\label{alg:square-well} The stable EDPD algorithm for event
    detection between two square-well particles $i$ and $j$.}
\end{algorithm}

\bibliography{logical-states}
\bibliographystyle{spmpsci}

\end{document}